\documentclass[traditabstract]{aa} 
\usepackage{graphicx}
\usepackage{txfonts}

\begin{document}

\title{Observational evidence for a broken Li Spite plateau and mass-dependent Li depletion
\thanks{Based in part on observations obtained at the W. M. Keck Observatory, 
the Nordic Optical Telescope on La Palma, and on data from 
the HIRES/Keck archive and the European Southern Observatory
ESO/ST-ECF Science Archive Facility}
}
\titlerunning{Li depletion in Spite plateau stars}

\newcommand{\teff}{$T_{\rm eff}$ }
\newcommand{\tsin}{$T_{\rm eff}$}

\author{
J. Mel\'endez\inst{1} \and
L. Casagrande\inst{2} \and
I. Ram{\'{\i}}rez\inst{2} \and
M. Asplund\inst{2} \and
W. J. Schuster\inst{3}
}


\institute{
Centro de Astrof\'{\i}sica da Universidade do Porto, Rua das Estrelas, 4150-762 Porto, Portugal  \and
Max Planck Institute for Astrophysics,
Postfach 1317, 85741 Garching, Germany \and
Observatorio Astron\'omico Nacional, Universidad Nacional Aut\'onoma de M\'exico,
Ensenada, B.C., CP 22800, Mexico
}

\date{Received ...; accepted ...}

\abstract{
We present NLTE Li abundances for 88 stars in the metallicity
range -3.5 $<$ [Fe/H]  $<$ -1.0. The effective temperatures are based on the infrared flux
method with improved $E(B-V)$ values obtained mostly from 
interstellar \ion{Na}{i}\,D lines. The Li abundances were derived through 
MARCS models and high-quality UVES+VLT, HIRES+Keck and FIES+NOT spectra,
and complemented with reliable equivalent widths from the literature.
The less-depleted stars with [Fe/H]$<-2.5$ and [Fe/H] $>-2.5$ fall into 
two well-defined plateaus 
of $A_{\rm Li}$ $= 2.18 \; (\sigma =0.04)$ and $A_{\rm Li}$  $=2.27 \; (\sigma =0.05)$, respectively.
We show that the two plateaus are flat, unlike previous
claims for a steep monotonic decrease in Li abundances with decreasing metallicities.
At all metallicities we uncover a fine-structure in the Li abundances of Spite
plateau stars, which we trace to Li depletion that depends on both metallicity and mass.
Models including atomic diffusion and turbulent mixing seem to reproduce the observed Li depletion 
assuming a primordial Li abundance $A_{\rm Li}$  = 2.64, which 
agrees well with current predictions ($A_{\rm Li}$ = 2.72) from standard
Big Bang nucleosynthesis. Adopting the Kurucz overshooting model atmospheres
increases the Li abundance by +0.08 dex to
$A_{\rm Li}$ = 2.72, which perfectly agrees with BBN+WMAP.
}


\maketitle

%

\section{Introduction}

One of the most important discoveries in the study of
the chemical composition of stars was made in 1982
by Monique and Fran\c cois Spite, who found an essentially constant
Li abundance in warm metal-poor stars (Spite \& Spite 1982), 
a result interpreted as a relic of
primordial nucleosynthesis. Due to its cosmological
significance, there have been many studies
devoted to Li in metal-poor field stars
(e.g., Ryan et al. 1999;
Mel\'endez \& Ram{\'{\i}}rez 2004, hereafter MR04; 
Boesgaard et al. 2005, hereafter B05;
Charbonnel \& Primas 2005; Asplund et al. 2006, hereafter A06;
Shi et al. 2007, hereafter S07; Bonifacio et al. 2007, hereafter B07;
Hosford et al. 2009; Aoki et al. 2009; Sbordone et al. 2010, hereafter S10),
with observed Li abundances at the lowest metallicities ([Fe/H] $\sim-$3)
from $A_{\rm Li}$  = 1.94 (B07) to $A_{\rm Li}$  = 2.37 (MR04).

A primordial Li abundance
of $A_{\rm Li}$  = 2.72$_{-0.06}^{+0.05}$ is predicted (Cyburt et al. 2008; 
see also Steigman 2009; Coc \& Vangioni 2010)
with the theory  of big bang nucleosynthesis (BBN) and
the baryon density obtained from WMAP data (Dunkley et al. 2009), 
which is a factor of 2-6 times higher than the Li abundance inferred
from halo stars.
There have been many theoretical studies on non-standard
BBN trying to explain the cosmological Li discrepancy by
exploring the frontiers of new physics 
(e.g. Coc et al. 2009; Iocco et al. 2009; 
Jedamzik \& Pospelov 2009; Kohri \& Santoso 2009). 
Alternatively, the Li problem 
could be explained by a reduction of the original Li stellar abundance
due to internal processes (i.e., by stellar depletion).
In particular, stellar models including atomic diffusion and mixing can 
deplete a significant fraction of the initial Li content (Richard et al. 2005; Piau 2008),
although such models 
depend on largely unconstrained free parameters.
On the other hand, it is not easy to reconcile the lack of 
observed abundance scatter in the Spite plateau
with substantial Li depletion (e.g. Ryan et al. 1999; A06).  
Due to the uncertainties in the Li abundances
and to the small samples available, only limited comparisons of 
models of Li depletion with stars in a broad range of mass and
metallicities  have been performed (e.g., Pinsonneault et al. 2002; B05).

The observed scatter in derived Li abundances in previous studies of
metal-poor stars can be as low as 0.03 dex (e.g. Ryan et al. 1999; A06),
fully consistent with the expected observational errors. Yet,
for faint stars observational errors as high as 0.2 dex have been reported (e.g., Aoki et al. 2009).
In order to provide meaningful comparisons with stellar depletion
models, precise Li abundances for a large sample
of stars  are needed. 
Here we present such a study for the first time 
for  a large sample 
of metal-poor stars (-3.5 $<$ [Fe/H] $<$ -1.0) with masses 
in a relatively broad mass range (0.6-0.9 M$_\odot$).

\section{Stellar sample and atmospheric parameters}

The sample was initially based on the 62 stars
analysed by MR04.
We added stars with published equivalent width (EW) 
measurements based on high-quality observations (Sect. 3), and included also 
stars for which we could obtain EW Li measurements
from UVES+VLT, HIRES+Keck and FIES+NOT spectra (either from our own
observations or from the archives). The resolving power of the
different instruments used ranges from $\sim$ 45,000 to 110,000, and the
S/N is typically $>$ 100. Some measurements taken from the literature
were obtained with lower resolving power ($\sim$35,000), or high resolving
power (R $\sim$ 50,000) but lower S/N. When necessary we averaged several
measurements in order to decrease the errors. 
The stars for which we could
not obtain EW better than $\sim$5\% 
were discarded, with a few exceptions (see Sect. 3).
Our adopted log $g$ values and metallicities [Fe/H] are mean
values (after discarding outliers) taken from 
the compilation of stellar parameters by Mel\'endez (in preparation),
who has updated the metallicity catalogue of Cayrel de Strobel et al. (2001)
from 4918 entries to more than 14000. 
We only keep stars with at least two previous spectroscopic 
analyses, so that their metallicities
and surface gravities have been confirmed at least by one other study. 
The sample has 88 stars with stellar parameters 
$5250 < T_{\rm eff} < 6600$ K,
$3.6 < \log g < 4.8$ and
$-3.5 < {\rm [Fe/H]} < -1.0$.

We obtained effective temperatures (\tsin) by a new
implementation (Casagrande et al. 2010, hereafter C10; see also Casagrande et al. 2006) 
of the infrared flux method (IRFM).
In our previous work on Li abundances in metal-poor stars (MR04),
\teff was obtained from the IRFM implementation 
by Ram{\'{\i}}rez \& Mel\'endez (2005), relying on the calibration of the bolometric 
correction by Alonso et al. (1995), which contains only a few low-metallicity stars;
thus, for very metal-poor turn-off stars, it is not well defined and may introduce systematic errors.
Indeed, C10 have found that below [Fe/H] $\simeq -2.5$, 
the Ram{\'{\i}}rez \& Mel\'endez (2005) \teff
go from being too cool by $\simeq100$\,K at [Fe/H]=$-$2.5 to 
too hot by $\simeq100$\,K at [Fe/H]=$-$3.5.

Our new IRFM implementation
determines the bolometric flux from observed broad-band photometry
(BVRIJHK), thus representing an important improvement 
over the \teff obtained by Ram{\'{\i}}rez \& Mel\'endez (2005).
Furthermore, the zero point of our \teff scale 
has been carefully checked for accuracy using solar twins (their
average \teff matching the solar one within a few K), spectral energy distributions 
measured with HST for a metal-rich star with planet and a metal-poor turn-off star, 
and stars with interferometric angular diameters
(for which our IRFM \teff and fluxes result in 
angular diameters only $0.6\pm1.7$\,\% smaller, corresponding to 
$18\pm50$\,K). Thus, contrary to most previous studies, the uncertainty of 
our absolute Li abundances is not dominated by the unknown 
zero point of the adopted \tsin\ scale.

In order to obtain \tsin\ values with IRFM, one needs good estimates of $E(B-V)$
to properly correct the photometric data. In our previous work we used
extinction maps (see Sect. 4 in Mel\'endez et al. 2006, hereafter M06), 
but in order to achieve a very high precision ($\sim$ 0.005 mag), 
we must rely on other methods. Thanks to the large radial velocity of
most metal-poor stars, the interstellar  \ion{Na}{i}\,D lines can be used to
obtain this precision. We employed this technique 
(Alves-Brito et al. 2010) when high-resolution spectra were available
($\sim$72\% of the cases), otherwise we used extinction maps (M06).
Our adopted $E(B-V)$ values are shown in Fig. 1 and have median errors of 0.003$\pm$0.002 mags,
implying errors in \teff of $\pm 15$\,K. Adding in errors in photometry,
bolometric flux, log $g$ and [Fe/H], we estimate a typical total internal error in \teff of $\pm 50$\,K.
\footnote{In some cases the \teff given in Table 1 may be slightly different
from the \teff given in C10 due to improved input
parameters (e.g., better E(B-V) values) in the IRFM used here. 
The conclusions of C10 are unaffected by these minor changes.}

\begin{figure}
\resizebox{\hsize}{!}{\includegraphics{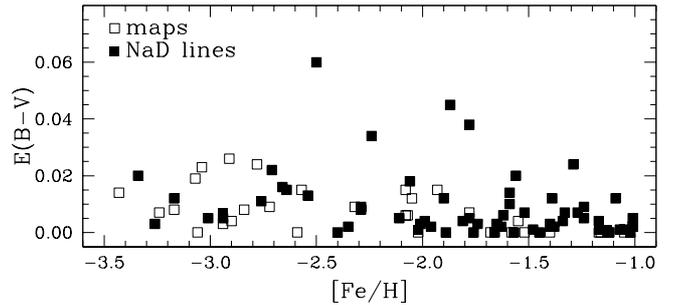}}
\caption{Adopted E(B-V) values vs. [Fe/H]. Filled squares are
values based on NaD lines and open squares on reddening maps.
}
\label{liteff}    
\end{figure}

We obtained stellar masses from the $\alpha$-enhanced $Y^2$ isochrones 
(Demarque et al. 2004). We interpolated a fine grid of models with
a step $\Delta$[Fe/H] = 0.02 dex and adopting [$\alpha$/Fe] = 0
for [Fe/H] $\geq$ 0, [$\alpha$/Fe] = $-$0.3 $\times$[Fe/H] for $-$1 $<$ [Fe/H] $<$ 0,
and [$\alpha$/Fe] = +0.3 for [Fe/H] $\leq$ $-$1. 
\footnote{The adopted relation between
[$\alpha$/Fe] and [Fe/H] ignores the detailed abundance patterns
of the two discs and halos (e.g., Ram{\'{\i}}rez et al. 2007, Nissen \& Schuster 2010).
}
At a given metallicity, we
searched for all solutions allowed by the error bars in \tsin,
Hipparcos parallaxes (reliably available for 2/3 of the sample) and [Fe/H],
adopting the median value in age and mass, 
and the standard deviation of the solutions is adopted as the error.
We also used the mean log $g$ values found in 
the literature to estimate another set of masses and ages. 
The masses obtained by both methods agree very well (median difference
[Hipparcos - literature] of $-$0.001 M$_\odot$).
The weighted averages of both results were adopted.
We find typical ages of $\sim$11 Gyr and masses in
the range of 0.6-0.9 M$_\odot$. In a few cases the adopted stellar parameters
resulted in low ages ($<$ 6 Gyr), probably due to errors in
log $g$. In those cases we estimated masses adopting an age=11 Gyr,
but the choice of age does not affect our conclusions
(e.g., $\Delta$age=+1 Gyr results in $\Delta$mass=$-$0.007 M$_\odot$ 
for a dwarf with [Fe/H]=$-$2 and \teff $\approx$6300 K).

\section{Li abundance analysis}
The LTE Li abundances were obtained with our code (A06)
with MARCS (Gustafsson et al. 2008) models and
the 2002 version of MOOG (Sneden 1973) with 
Kurucz overshooting (Castelli et al. 1997) models 
(which have improved overshooting with respect to earlier models
that showed several problems described in van't Veer-Menneret 
\& Megessier 1996).
Both model atmospheres give similar Li abundances
except for a zero-point difference of $\sim$0.08 dex (Kurucz overshooting $-$ MARCS).
According to MR04 the difference in $A_{\rm Li}$ 
between Kurucz overshooting and no-overshooting models is also
$\sim$0.08 dex, so Kurucz no-overshooting models give
similar $A_{\rm Li}$  as the MARCS models.
The results presented herein will be based on the MARCS models,
because the overshoot option implemented by Kurucz 
is not supported by 3D hydrodynamical model atmospheres
(Asplund et al. 1999; see also Heiter et al. 2002);
we will nevertheless mention the implication of using Kurucz overshooting models
for completeness.
We adopt the 1D non-LTE corrections by 
Lind et al. (2009a), which are based on the most recent radiative and 
collisional data (Barklem et al. 2003). 
The difference between LTE and non-LTE is relatively constant 
for our stars so that similar conclusions
would have been obtained assuming LTE,
yet, for an accurate analysis NLTE must be taken 
into account because it affects the absolute level of the 
Spite plateau. For metal-poor ([Fe/H] $< -$1) stars 
with \teff $>$ 6000 K the typical NLTE correction is
$-0.06\pm0.01$ dex.

We measured the EW of the Li feature at 6707.8 \AA\
using UVES+VLT, HIRES+Keck and FIES+NOT high-resolution spectra, from which
we obtain typical errors in EW of 0.7 m\AA, which includes
uncertainties in continuum placement.
We complemented our measurements with data from the literature
(B05; A06; B07; S07; Asplund \& Mel\'endez 2008, hereafter AM08; S10).
An important improvement with respect to our previous work 
is that now we select stars
with errors in EW below 5\% (typically 
$\sim$ 2-3\%), 
instead of the 10\% limit adopted in 
MR04. The only
exceptions are the cool dwarfs HD64090 and BD+38 4955, 
which are severely depleted in Li and have EW errors 
of 8\% and 10\%, respectively, and the
very metal-poor stars (B07)
BPS CS29518-0020 (5.2\%) and BPS CS29518-0043 
(6.4\%), which were kept due to their
low metallicity. 

The main sources of error are the uncertainties in equivalent widths
and \tsin, which in our work have typical values of only 2.3\% and
50 K, implying abundance errors of 0.010 dex and 0.034 dex, respectively,
and a total error in $A_{\rm Li}$  of $\sim$ 0.035 dex. This
low error in $A_{\rm Li}$  is confirmed by the star-to-star scatter of the
Li plateau stars, which have similar low values (e.g. $\sigma$ = 0.036 dex
for [Fe/H] $< -2.5$, see below). 
Our Li abundances and stellar parameters are given in Table 1 (online material).

\begin{figure}
\resizebox{\hsize}{!}{\includegraphics{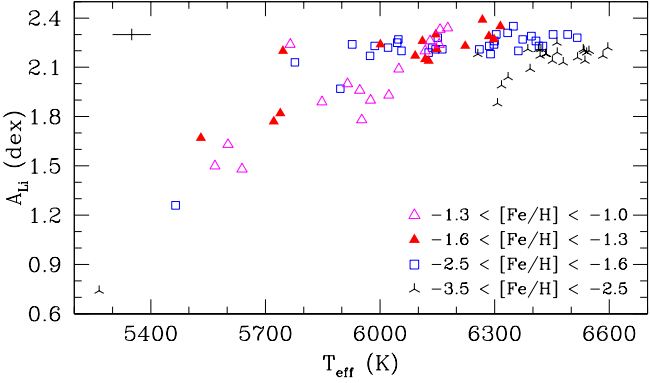}}
\caption{Li abundances vs. \teff for our sample stars in different
metallicity ranges. A typical error bar is shown. 
}
\label{ebv}    
\end{figure}

\begin{figure}
\resizebox{\hsize}{!}{\includegraphics{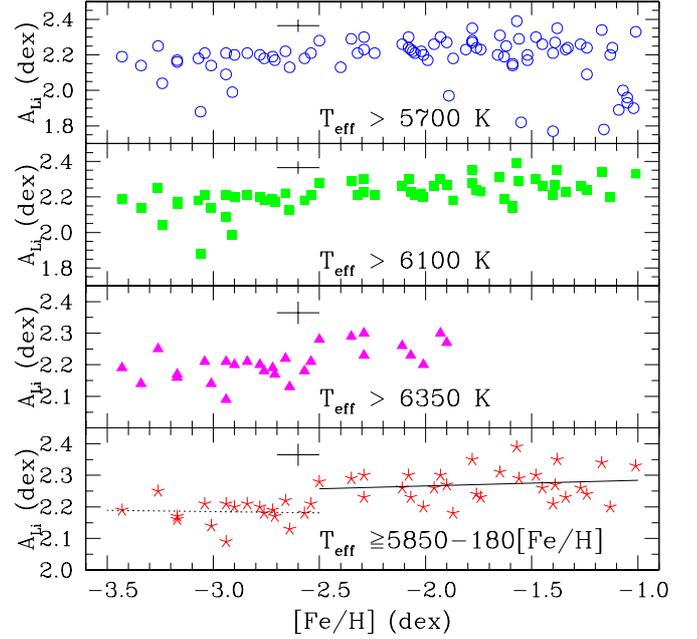}}
\caption{Li abundances for stars with 
\teff $>$ 5700 K (open circles), $>$ 6100 K (filled
squares), $>$ 6350 K (filled triangles) and 
$\geq$ 5850 - 180$\times$[Fe/H] (stars). 
In the bottom panel stars above the cutoff in \teff
fall into two flat plateaus with $\sigma$=0.04 and 0.05 dex 
for [Fe/H] $<-2.5$ (dotted line) 
and [Fe/H] $\geq -2.5$ (solid line), respectively. 
}
\label{lifeh}    
\end{figure}

\section{Discussion}

\subsection{The \teff cutoff of the Spite plateau}

Despite the fact that Li depletion depends on mass 
(e.g. Pinsonneault et al. 1992), 
this variable has been ignored by most previous studies. Usually
a cutoff in \teff is imposed to exclude severely Li-depleted stars in the Spite plateau,
with a wide range of adopted cutoffs, such as 5500 K (Spite \& Spite 1982), 5700 K (B05),
6000 K (MR04; S07) and $\sim$6100 K for stars with [Fe/H] $< -2.5$ (Hosford et al. 2009).

At a given mass, the \teff of metal-poor stars has a strong metallicity-dependence 
(e.g. Demarque et al. 2004). As shown
in Figs. 11-12 of M06, the \teff of turnoff stars increases for
decreasing metallicities.
Hence, a metallicity-independent cutoff in \teff
may be an inadequate way to exclude low-mass Li-depleted stars
from the Spite plateau. As show in Fig. 2, where $A_{\rm Li}$ in different metallicity bins
is shown as a function of \tsin, stars with lower \teff in a given metallicity
regime are typically the stars with the lowest Li abundances, 
an effect that can be seen even in the sample
stars with the lowest metallicities ([Fe/H] $\sim -3$). This is ultimately
so because the coolest stars are typically the least massive,
and therefore have been more depleted in Li (see Sect. 4.3). 

In Fig. 3 we show the Li abundance for cutoffs = 5700 K (open circles),
6100 K (filled squares) and 6350 K (filled triangles). Using a hotter
cutoff is useful to eliminate the most Li-depleted stars at low
metallicities, but it removes from the Spite plateau stars
with [Fe/H] $>$ -2.
Imposing a hotter \teff cutoff at low 
metallicities and a cooler cutoff at high metallicities
eliminates the most Li-depleted stars at low metallicities,
but keeps the most metal-rich stars in the Spite plateau. 
We propose such a metallicity-dependent cutoff below.

\subsection{Two flat Spite plateaus}

Giving the shortcomings of a constant \teff cutoff,
we propose an empirical cutoff of \teff $=$ 5850 - 180$\times$[Fe/H].
The stars above this cutoff are shown as stars in the bottom panel of Fig. 3.
Our empirical cutoff excludes only the 
most severely Li-depleted stars, i.e., the stars that remain
in the Spite plateau may still be affected by depletion. 
The less Li-depleted stars in the bottom panel of Fig. 3 show two 
well-defined groups separated at [Fe/H] $\sim-2.5$
(as shown below, this break represents a real discontinuity),
which have essentially zero slopes (within the error bars) and very low star-to-star
scatter in their Li abundances.
The first group 
has $-2.5 \leq$ [Fe/H] $< -1.0$ and $<$A$_{\rm Li}$$>_1$ = 2.272 ($\sigma$=0.051) dex
and a slope of 0.018$\pm$0.026, i.e., 
flat within the uncertainties. 
The second group is more metal poor ([Fe/H] $< -2.5)$
and has $<$A$_{\rm Li}$$>_2$ = 2.184 dex ($\sigma = 0.036$) dex.
The slope of this second group is also zero ($-$0.008$\pm$0.037).
Adopting a more conservative exponential cutoff
obtained from $Y^2$ isochrones (Demarque et al. 2004),
which for a 0.79 M$_\odot$ star can be fit by
\teff = 6698 $-$2173 $\times$ $e^{\rm [Fe/H]/1.021}$, we
would also recover a flat Spite plateau, although only
stars with [Fe/H] $> -$2.5 are left using this more restrictive
cut-off. Thus, the flatness of the Spite plateau is
independent of applying a linear or an exponential cutoff.

Adopting a constant cutoff in \teff we also find flat plateaus.
For example adopting a cutoff of \teff $>$ 6100 K (filled squares in Fig. 3) we find 
in the most metal-rich plateau ([Fe/H] $\geq -2.5$) 
no trend between Li and [Fe/H] (slope = 0.019$\pm$0.025, 
Spearman rank correlation coefficient $r_{\rm Spearman}$=0.1 and a probability of
0.48 (i.e., 48\%) of a correlation arising by pure chance for [Fe/H] $\geq -$2.5), 
while for the most metal-poor plateau ([Fe/H] $< -$2.5) we
also do not find any trend within the errors (slope = 0.058$\pm$0.072,
$r_{\rm Spearman}$=0.2 and 41\% probability of a spurious correlation). Using a
hotter cutoff (\teff $>$ 6350 K, filled triangles in Fig. 3) we obtain also two flat plateaus with
slope = -0.040$\pm$0.063 ($r_{\rm Spearman}$ = -0.2, probability = 60\%) for [Fe/H] $\geq -$2.5 
and slope = 0.008$\pm$0.035 ($r_{\rm Spearman}$ = 0.1, probability = 68\%) for [Fe/H] $< -$2.5.

\begin{figure}
\resizebox{\hsize}{!}{\includegraphics{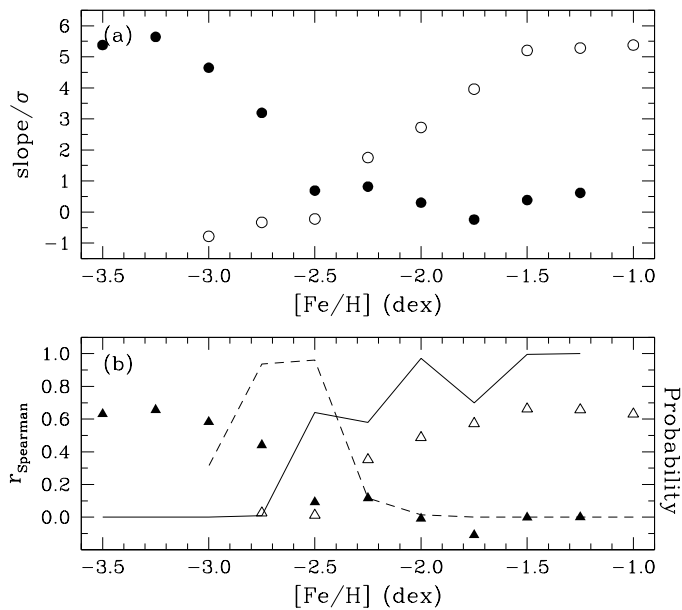}}
\caption{(a) filled circles show the slope/$\sigma$
of $A_{\rm Li}$ vs. [Fe/H] in the range
X$_{\rm min}$ $<$ [Fe/H] $<-1.0$, where X$_{\rm min}$ 
is in the interval [-3.5,-1.25]. For all values of X$_{\rm min}$ $\geq -2.5$, the
slope is insignificant ($<<$ 3$\sigma$); only when 
stars with [Fe/H] $< -2.5$ are included, a slope 
is forced in $A_{\rm Li}$ vs. [Fe/H].
Open circles show the slope/$\sigma$ for the range
$-3.25<$ [Fe/H] $<$ X$_{\rm max}$, where $-$3.0 $\leq$ X$_{\rm max}$ $\leq -$1.0.
For X$_{\rm max}$ $< -2.5$ 
the slope is negligible ($<<$ 3$\sigma$),
but when stars with [Fe/H] $\geq -2.5$ are included,
a slope is generated.
(b) The correlation coefficient $r_{\rm Spearman}$
is shown by triangles, with filled and open symbols with similar 
meaning as in panel (a). The probabilities of a correlation 
between $A_{\rm Li}$ and [Fe/H] by pure chance, associated to the 
filled and open triangles,
are shown by solid and dashed lines, respectively. 
No meaningful correlation (probability $<<$ 1) 
exists for either [Fe/H] $< -$2.5 or [Fe/H] $\geq -$2.5.
}
\label{libreak}
\end{figure}

In Fig. 4 we demonstrate that the break at [Fe/H] $\sim -2.5$ is statistically significant.
In panel (a) the slope/$\sigma$
in the $A_{\rm Li}$ vs. [Fe/H] plot for the range
X$_{\rm min}$ $<$ [Fe/H] $<-1.0$  are shown as filled circles, where X$_{\rm min}$ 
varies within [-3.50,-1.25]. For X$_{\rm min}$ $\geq -2.5$, the
slope is insignificant ($<<$ 3$\sigma$), and only when 
stars with [Fe/H] $< -2.5$ are included a measurable slope 
is forced in the  $A_{\rm Li}$ vs. [Fe/H] relation.
The opposite test is shown by open circles,
where we show slope/$\sigma$ for the range
$-$3.25 $<$ [Fe/H] $<$ X$_{\rm max}$, where X$_{\rm max}$ changes
from [$-$3.00,$-$1.00]. 
For X$_{\rm max}$ $< -2.5$ the slope is negligible ($<<$ 3$\sigma$), 
and only when stars with [Fe/H] $\geq -2.5$ are included,
a slope is produced between $A_{\rm Li}$ and [Fe/H].
The correlation coefficient $r_{\rm Spearman}$ and the probability of
a correlation between $A_{\rm Li}$ and [Fe/H] by 
pure chance are shown in panel (b). 
Again, this plot shows that no correlation between $A_{\rm Li}$ and [Fe/H] 
exists {\it within} the two groups ($-3.5<$ [Fe/H] $<-2.5$ and $-2.5\geq$ [Fe/H] $\geq-1.0$),
and that only when stars {\it between} the two groups are mixed,
significant (probability $\sim$ 0) correlations of $A_{\rm Li}$ with [Fe/H] are generated.
Systematically lower E(B-V) values (by $\sim$0.03 mags) in the
 most metal-poor plateau ([Fe/H]  $< -$2.5) could produce a more Li-depleted plateau.
However, those E(B-V) values (Fig.~1)
are actually slightly higher ($\sim$0.007 mags, i.e., $\sim$33 K) than those for [Fe/H]  $\geq -$2.5
(mainly due to a high number of unreddened nearby more metal-rich stars),
thus not explaining the existence of two different plateaus.

Previous claims of a steep monotonic decrease in Li abundance with 
decreasing metallicity 
(e.g. Ryan et al. 1999; A06; B07; Hosford et al. 2009)
are probably due to the mix of stars from the two
different groups,
forcing a monotonic dependence with metallicity.
Our large sample of homogeneous and precise Li abundances 
that covers a broad metallicity range ($-3.5 <$ [Fe/H] $< -1.0$)
does not support these claims. Nevertheless, a hint of two different
groups in the Spite plateau was already found by A06,
who found a change in the slope of the Spite plateau at [Fe/H] $\approx$ $-$2.2.
Also, in the combined A06+B07 sample (Fig. 7 of B07),
there are two different groups: stars with [Fe/H] $\gtrsim -2.6$ have $A_{\rm Li}$ $>$ 2.2,
while stars with [Fe/H] $\lesssim -2.6$ have $A_{\rm Li}$ $<$ 2.2. 
Although in the study by MR04 a flat Spite plateau is found
in the range $-3.4 <$ [Fe/H] $< -1$, this is due to the overestimation
of \teff below [Fe/H] $< -2.5$, thus overestimating $A_{\rm Li}$ at
low metallicities and forcing a flat plateau from [Fe/H] = $-3.4$ to $-$1.

\subsection{Correlation between Li and mass}

Models of Li depletion predict that the least massive stars are
the most depleted in Li, but due to the limitations of previous
samples, these predictions have not been thoroughly tested at
different metallicity regimes in metal-poor stars.\footnote{Except
for the work of B05, who provide comparisons 
at different metallicities but for cool severely Li-depleted dwarfs,
i.e. probably of lower mass than most stars shown in Fig. 5}
In Fig. 5 we show our Li abundances 
as a function of stellar mass for different metallicity ranges.
As can be seen, the Li plateau stars have a clear dependence with mass for all
metallicity regimes. Excluding stars with mass $<$ 0.7M$_\odot$ (including those
stars will result in even stronger correlations), linear fits result in
slopes of 6, 3, 2, 2 dex M$_\odot^{-1}$ 
for stars in the metallicity ranges $-1.3 <$ [Fe/H] $< -1.0$, 
$-1.6 <$ [Fe/H] $< -1.3$, $-2.5 <$ [Fe/H] $< -1.6$, and $-3.5 <$ [Fe/H] $< -2.5$.
The slopes are significant at the 8, 2, 5, 1 $\sigma$ level, respectively.
The correlation coefficient $r_{\rm Spearman}$ is 0.9, 0.6, 0.6, 0.3, and
the probability of a correlation between  $A_{\rm Li}$ and mass 
arising by pure chance is very small: 5$\times 10^{-5}$, 
3$\times 10^{-2}$, 1$\times 10^{-3}$, and 1.3$\times 10^{-1}$,
for stars in the same metallicity ranges as above. Thus, the correlations of $A_{\rm Li}$ and mass
in different metallicity regimes are very significant.

Recently, Gonz\'alez Hern\'andez et al. (2008, hereafter G08) have studied the metal-poor 
([Fe/H] $\sim-$3.5) double-lined spectroscopic binary BPS CS22876-032,
providing thus crucial data to test our Li-mass trend.
With our method and the stellar parameters of G08,
we obtain a mass ratio of 0.89, very close to their value (0.911$\pm$0.022)
obtained from an orbital solution.
For the primary we obtain $M_A$ = 0.776 M$_\odot$, and adopting
the mass ratio of G08, $M_B$ = 0.707 M$_\odot$ is obtained for the secondary. 
The LTE Li abundances were taken from
G08 and corrected for NLTE effects ($\sim-$0.05 dex). 
The components of the binary are shown as circles in Fig. 5,
nicely following the trend of the most metal-poor stars. 
Including this binary
in our sample would strengthen the Li-mass correlation of
stars with [Fe/H] $< -$2.5. A slope = 3 dex  M$_\odot^{-1}$ 
significant at the 3$\sigma$ level is obtained,
with $r_{\rm Spearman}$ = 0.5 and a low probability 
(2$\times 10^{-2}$) of the trend being spurious.

While mass-dependent Li depletion
is expected from standard models of stellar evolution,
this should only occur at significantly lower masses than considered here.
These stellar models only predict very minor $^7$Li depletion ($\la 0.02$\,dex)
for metal-poor turn-off stars (e.g. Pinsonneault et al. 1992), 
which is far from sufficient to explain the $\sim$0.5\,dex discrepancy
between the observed Li abundance and predictions from BBN+WMAP.
Bridging this gap 
would thus require invoking 
stellar models that include additional processes normally not
accounted for, such as rotationally-induced mixing or diffusion. 
In Fig. 5 we confront the predictions of Richard et al. (2005) with our inferred
stellar masses and Li abundances. 
The models include the effects of atomic diffusion, radiative acceleration, and gravitational settling,
but moderated by a parameterized turbulent mixing (T6.0, T6.09, and T6.25, 
where higher numbers mean higher turbulence);
so far only the predictions for [Fe/H]=$-$2.3 are available
for different turbulent mixing models.  
The agreement is very good when adopting a turbulent model of T6.25 
(see Richard et al. for the meaning of this notation)  
and an initial $A_{\rm Li}  = 2.64$; 
had the Kurucz convective overshooting
models been adopted, the required initial abundance to explain our observational data 
would correspond to $A_{\rm Li} = 2.72$.
Two other weaker turbulence models that produce smaller overall Li depletions 
are also shown in Fig. 5, but they are less
successful in reproducing the observed Li abundance pattern.
Our best-fitting turbulence model (T6.25) is different
from that (T6.0) required to explain the abundance pattern in the globular cluster
NGC\,6397 at similar metallicity (Korn et al. 2006; Lind et al. 2009b).
Another problem with adopting this high turbulence is that the expected corresponding
$^6$Li depletion would amount to $>1.6$\,dex and thus imply an initial
$^6$Li abundance at least as high as the primordial $^7$Li abundance if
the claimed $^6$Li detections in some halo stars (A06; AM08) are real.

Our results imply that the Li abundances observed in Li plateau
stars have been depleted from their original values and therefore
do not represent the primordial Li abundance.
It appears that the observed Li abundances in metal-poor stars
can be reasonably well reconciled with the predictions from standard Big Bang nucleosynthesis
(e.g. Cyburt et al. 2008) by means of stellar evolution models that
include Li depletion through diffusion and turbulent mixing
(Richard et al. 2005). 
We caution however that although encouraging, our results 
should not be viewed as proof of the 
Richard et al. models until the
free parameters required for the stellar modelling are 
better understood from physical principles.

\begin{figure}
\resizebox{\hsize}{!}{\includegraphics{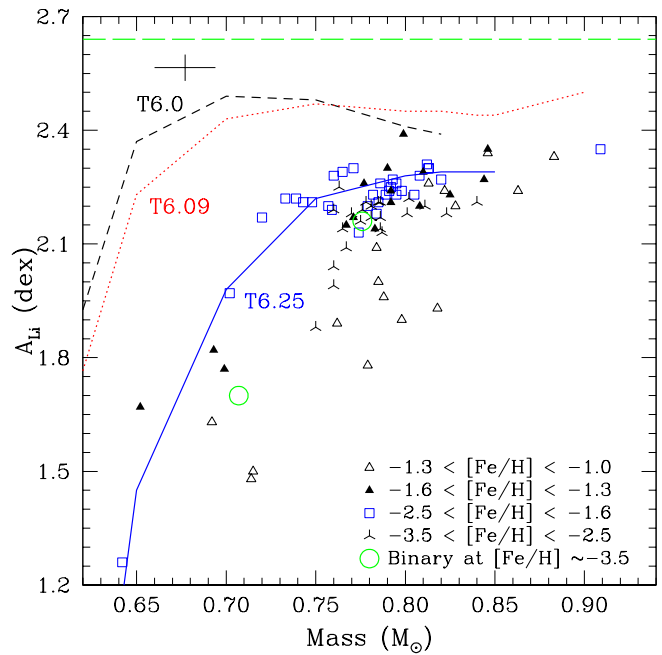}}
\caption{Li abundances as a function of stellar mass in different
metallicity ranges. The metal-poor ([Fe/H] $\sim-$3.5) binary 
BPS CS22876-032 (G08) is represented by open circles, nicely fitting the Li-mass trend.
Models at [Fe/H] = $-2.3$ including diffusion and
T6.0 (short dashed line), T6.09 (dotted line) 
and T6.25 (solid line) turbulence (Richard et al. 2005) are shown. 
The models were rescaled to an initial $A_{\rm Li}$=2.64 (long dashed line) 
and by $\Delta M = +0.05$\,M$_\odot$.
}
\label{massli}    
\end{figure}

\begin{acknowledgements}
We thank the referee for valuable comments to strengthen the
presentation of our results.
We thank J. R. Shi for providing the EW measurements
used in S07, O. Richard for providing tables with
the Li depletion models, R. Collet, P. Nissen, A. Garc\'{i}a Per\'ez, 
D. Fabbian and L. Sbordone for providing data 
to estimate E(B-V) from interstellar \ion{Na}{i}\,D lines, 
K.Lind for providing NLTE abundance corrections in electronic form.
This work has been partially supported by FCT (project PTDC/CTE-AST/65971/2006, and Ciencia 2007 program).
\end{acknowledgements}

\Online

\scriptsize
\begin{table}
\caption{Parameters and Li abundances of our metal-poor stars}
\label{tab1}
\centering 
\renewcommand{\footnoterule}{}  
\begin{tabular}{lllllllllllllllllllllllllllllllllllllllllllllll} 
\hline
\hline 
{Star}    &  EW$\pm\sigma$ & ref. & {E(B-V)$\pm\sigma$} & ref. & \teff & log $g$ & [Fe/H] &  mass$\pm\sigma$ & $A_{\rm Li}^{\rm LTE}$ & $A_{\rm Li}^{\rm NLTE}$ \\
{     }      & (m\AA)     &    & (mag)              & {}     &  (K)  & (dex)   & (dex)   & (M$_\odot)$     & (dex) & (dex) \\
\hline
BD+01 3597        & 24.9$\pm$0.5 & A06         & 0.045$\pm$0.005 &NaD       & 6289 & 4.04 & -1.87 & 0.784$\pm$0.018 & 2.22 & 2.18 \\
BD+02 3375        & 33.3$\pm$1.0 & MR04+S07    & 0.034$\pm$0.010 &NaD       & 6163 & 4.13 & -2.24 & 0.748$\pm$0.010 & 2.27 & 2.21 \\
BD+02 4651        & 31.6$\pm$0.8 & UVES archive& 0.038$\pm$0.004 &NaD       & 6349 & 3.79 & -1.78 & 0.909$\pm$0.062 & 2.42 & 2.35 \\
BD+03 0740        & 20.8$\pm$0.5 & A06         & 0.022$\pm$0.003 &NaD       & 6419 & 3.97 & -2.71 & 0.781$\pm$0.021 & 2.21 & 2.17 \\
BD+09 0352        & 34.0$\pm$1.5 & MR04        & 0.012$\pm$0.004 &maps      & 6145 & 4.32 & -2.05 & 0.743$\pm$0.015 & 2.28 & 2.21 \\
BD+09 2190        & 16.6$\pm$0.5 & A06         & 0.015$\pm$0.003 &NaD       & 6479 & 4.01 & -2.64 & 0.787$\pm$0.030 & 2.18 & 2.13 \\
BD+11 0468        & 26.0$\pm$1.2 & B05         & 0.004$\pm$0.004 &maps      & 5739 & 4.58 & -1.55 & 0.693$\pm$0.015 & 1.85 & 1.82 \\
BD+17 4708        & 27.6$\pm$0.5 & A06         & 0.010$\pm$0.002 &NaD       & 6127 & 4.04 & -1.59 & 0.783$\pm$0.014 & 2.19 & 2.14 \\
BD+24 1676        & 24.2$\pm$1.1 & UVES archive& 0.013$\pm$0.004 &NaD       & 6387 & 3.84 & -2.54 & 0.840$\pm$0.071 & 2.27 & 2.21 \\
BD+26 4251        & 35.5$\pm$0.4 & UVES (NS10) & 0.007$\pm$0.003 &NaD       & 6131 & 4.39 & -1.27 & 0.813$\pm$0.014 & 2.33 & 2.26 \\
BD+29 2091        & 37.1$\pm$1.0 & UVES archive& 0.004$\pm$0.003 &NaD       & 5974 & 4.58 & -1.99 & 0.720$\pm$0.014 & 2.21 & 2.17 \\
BD+34 2476        & 23.8$\pm$1.0 & MR04+B05+S07& 0.006$\pm$0.005 &maps      & 6416 & 3.95 & -2.07 & 0.805$\pm$0.028 & 2.30 & 2.23 \\
BD+36 2165        & 33.1$\pm$0.9 & FIES/NOT    & 0.002$\pm$0.002 &NaD       & 6315 & 4.28 & -1.38 & 0.846$\pm$0.011 & 2.42 & 2.35 \\
BD+38 4955        &  4.9$\pm$0.5 & MR04        & 0.000$\pm$0.010 &nearby    & 5265 & 4.60 & -2.59 & 0.578$\pm$0.010 & 0.75 & 0.74 \\
BD+42 2667        & 31.7$\pm$0.5 &HIRES archive& 0.003$\pm$0.001 &NaD       & 6148 & 4.26 & -1.40 & 0.792$\pm$0.010 & 2.27 & 2.21 \\
BD+42 3607        & 39.5$\pm$0.5 &HIRES archive& 0.018$\pm$0.003 &NaD       & 6021 & 4.59 & -2.06 & 0.733$\pm$0.015 & 2.29 & 2.22 \\
BD+51 1696        & 24.0$\pm$0.6 & FIES/NOT    & 0.000$\pm$0.002 &NaD+nearby& 5722 & 4.59 & -1.40 & 0.699$\pm$0.010 & 1.80 & 1.77 \\
BD-04 3208        & 23.9$\pm$0.5 & A06         & 0.009$\pm$0.003 &NaD       & 6491 & 3.98 & -2.29 & 0.813$\pm$0.021 & 2.38 & 2.30 \\
BD-10 0388        & 29.1$\pm$0.5 & A06         & 0.009$\pm$0.003 &maps      & 6260 & 3.98 & -2.32 & 0.785$\pm$0.018 & 2.27 & 2.21 \\
BD-13 3442        & 20.4$\pm$0.5 & A06         & 0.011$\pm$0.002 &NaD       & 6435 & 3.93 & -2.76 & 0.801$\pm$0.040 & 2.22 & 2.18 \\
CD-24 17504       & 18.2$\pm$0.5 & HIRES       & 0.020$\pm$0.005 &NaD       & 6451 & 4.13 & -3.34 & 0.765$\pm$0.019 & 2.18 & 2.14 \\
CD-30 18140       & 27.6$\pm$0.5 & A06         & 0.012$\pm$0.003 &NaD       & 6373 & 4.13 & -1.90 & 0.793$\pm$0.010 & 2.35 & 2.27 \\
CD-33 01173       & 16.0$\pm$0.5 & A06         & 0.005$\pm$0.005 &NaD       & 6536 & 4.29 & -3.01 & 0.786$\pm$0.016 & 2.18 & 2.14 \\
CD-33 03337       & 39.0$\pm$0.5 & A06         & 0.007$\pm$0.006 &NaD       & 6001 & 4.00 & -1.33 & 0.792$\pm$0.010 & 2.30 & 2.24 \\
CD-35 14849       & 28.4$\pm$0.5 & A06         & 0.002$\pm$0.002 &NaD       & 6396 & 4.22 & -2.35 & 0.765$\pm$0.010 & 2.37 & 2.29 \\
CD-48 02445       & 25.6$\pm$0.5 & A06         & 0.015$\pm$0.008 &maps      & 6453 & 4.25 & -1.93 & 0.813$\pm$0.016 & 2.38 & 2.30 \\
G011-044          & 33.0$\pm$1.2 & B05         & 0.006$\pm$0.006 &maps      & 6304 & 4.30 & -2.08 & 0.771$\pm$0.015 & 2.38 & 2.30 \\
G024-003          & 28.1$\pm$1.3 & UVES archive& 0.014$\pm$0.004 &NaD       & 6118 & 4.27 & -1.59 & 0.767$\pm$0.012 & 2.19 & 2.15 \\
G053-041          & 26.3$\pm$0.5 & FIES/NOT    & 0.009$\pm$0.002 &NaD       & 6049 & 4.31 & -1.24 & 0.784$\pm$0.012 & 2.14 & 2.09 \\
G064-012          & 23.7$\pm$0.5 & HIRES (AM08)& 0.003$\pm$0.002 &NaD       & 6463 & 4.17 & -3.26 & 0.763$\pm$0.010 & 2.32 & 2.25 \\
G064-037          & 16.4$\pm$0.5 & HIRES (AM08)& 0.012$\pm$0.003 &NaD       & 6583 & 4.20 & -3.17 & 0.786$\pm$0.012 & 2.21 & 2.17 \\
G075-031          & 37.8$\pm$0.5 & A06         & 0.005$\pm$0.001 &NaD       & 6157 & 4.19 & -1.01 & 0.883$\pm$0.015 & 2.40 & 2.33 \\
G114-042          & 25.2$\pm$1.2 & UVES (NS10) & 0.012$\pm$0.002 &NaD       & 5848 & 4.40 & -1.09 & 0.762$\pm$0.015 & 1.91 & 1.89 \\
G192-043          & 29.6$\pm$1.4 & FIES/NOT    & 0.012$\pm$0.003 &NaD       & 6298 & 4.39 & -1.39 & 0.844$\pm$0.017 & 2.35 & 2.27 \\
HD003567          & 38.3$\pm$0.5 & A06         & 0.004$\pm$0.003 &NaD       & 6177 & 4.14 & -1.17 & 0.846$\pm$0.010 & 2.41 & 2.34 \\
HD016031          & 28.4$\pm$0.6 & UVES archive& 0.003$\pm$0.003 &NaD       & 6286 & 4.17 & -1.74 & 0.789$\pm$0.011 & 2.29 & 2.23 \\
HD019445          & 34.9$\pm$0.5 & A06 + HIRES & 0.000$\pm$0.000 &NaD+nearby& 6136 & 4.45 & -2.02 & 0.739$\pm$0.011 & 2.29 & 2.22 \\
HD024289          & 47.6$\pm$1.7 & MR04        & 0.015$\pm$0.006 &maps      & 5927 & 3.73 & -2.08 & 0.798$\pm$0.054 & 2.31 & 2.24 \\
HD029907          & 26.0$\pm$1.3 & MR04        & 0.000$\pm$0.000 &nearby    & 5531 & 4.63 & -1.58 & 0.652$\pm$0.010 & 1.69 & 1.67 \\
HD031128          & 30.9$\pm$1.1 & UVES archive& 0.000$\pm$0.000 &nearby    & 6092 & 4.51 & -1.52 & 0.771$\pm$0.011 & 2.22 & 2.17 \\
HD034328          & 35.8$\pm$1.0 & UVES archive& 0.000$\pm$0.000 &NaD+nearby& 6056 & 4.50 & -1.66 & 0.757$\pm$0.012 & 2.27 & 2.20 \\
HD059392          & 39.2$\pm$0.5 & A06         & 0.006$\pm$0.003 &NaD       & 6045 & 3.87 & -1.62 & 0.792$\pm$0.012 & 2.32 & 2.25 \\
HD064090          & 12.1$\pm$1.0 & MR04        & 0.000$\pm$0.000 &nearby    & 5465 & 4.61 & -1.68 & 0.642$\pm$0.010 & 1.27 & 1.26 \\
HD074000          & 24.0$\pm$1.0 &UVES arc+MR04& 0.003$\pm$0.002 &NaD       & 6362 & 4.12 & -2.01 & 0.779$\pm$0.010 & 2.25 & 2.20 \\
HD084937          & 25.3$\pm$0.5 & HIRES (AM08)& 0.005$\pm$0.002 &NaD       & 6408 & 3.93 & -2.11 & 0.786$\pm$0.010 & 2.32 & 2.26 \\
HD094028          & 36.9$\pm$0.9 & UVES archive& 0.000$\pm$0.000 &NaD+nearby& 6111 & 4.36 & -1.45 & 0.777$\pm$0.010 & 2.33 & 2.26 \\
HD102200          & 33.0$\pm$0.5 & A06         & 0.005$\pm$0.004 &NaD       & 6155 & 4.20 & -1.24 & 0.822$\pm$0.010 & 2.30 & 2.24 \\
HD108177          & 31.2$\pm$0.9 & UVES archive& 0.003$\pm$0.002 &NaD       & 6334 & 4.41 & -1.65 & 0.812$\pm$0.010 & 2.40 & 2.31 \\
HD116064          & 29.7$\pm$0.6 & UVES archive& 0.000$\pm$0.003 &NaD+nearby& 5896 & 4.41 & -1.89 & 0.702$\pm$0.010 & 2.03 & 1.97 \\
HD122196          & 40.7$\pm$0.6 & UVES archive& 0.004$\pm$0.002 &NaD       & 5986 & 3.73 & -1.81 & 0.782$\pm$0.010 & 2.29 & 2.23 \\
HD126681          & 14.0$\pm$0.4 & UVES (NS10) & 0.000$\pm$0.001 &NaD+nearby& 5639 & 4.57 & -1.17 & 0.714$\pm$0.013 & 1.48 & 1.48 \\
HD132475          & 56.6$\pm$0.5 & UVES archive& 0.007$\pm$0.003 &NaD       & 5746 & 3.78 & -1.52 & 0.808$\pm$0.015 & 2.25 & 2.20 \\
HD140283          & 47.5$\pm$0.5 & A06+AM08    & 0.000$\pm$0.002 &NaD+nearby& 5777 & 3.62 & -2.40 & 0.774$\pm$0.010 & 2.17 & 2.13 \\
HD160617          & 40.5$\pm$0.5 & A06         & 0.005$\pm$0.004 &NaD       & 6048 & 3.73 & -1.78 & 0.820$\pm$0.023 & 2.33 & 2.27 \\
HD163810          & 21.6$\pm$0.7 & UVES (NS10) & 0.024$\pm$0.006 &NaD       & 5602 & 4.59 & -1.29 & 0.692$\pm$0.013 & 1.64 & 1.63 \\
HD166913          & 38.7$\pm$0.6 & UVES archive& 0.000$\pm$0.000 &NaD+nearby& 6268 & 4.29 & -1.57 & 0.799$\pm$0.010 & 2.46 & 2.39 \\
HD181743          & 38.0$\pm$1.4 & MR04        & 0.007$\pm$0.007 &M06       & 6151 & 4.48 & -1.78 & 0.760$\pm$0.013 & 2.36 & 2.28 \\
HD189558          & 58.2$\pm$0.7 & UVES archive& 0.000$\pm$0.004 &NaD+nearby& 5765 & 3.78 & -1.12 & 0.863$\pm$0.014 & 2.29 & 2.24 \\
HD193901          & 28.0$\pm$0.6 & UVES archive& 0.001$\pm$0.001 &NaD       & 5915 & 4.51 & -1.07 & 0.785$\pm$0.014 & 2.05 & 2.00 \\
HD194598          & 31.5$\pm$0.5 &     HIRES   & 0.000$\pm$0.000 &NaD+nearby& 6118 & 4.37 & -1.13 & 0.828$\pm$0.010 & 2.26 & 2.20 \\
HD199289          & 19.7$\pm$0.8 & UVES archive& 0.000$\pm$0.000 &NaD+nearby& 5975 & 4.35 & -1.02 & 0.798$\pm$0.010 & 1.93 & 1.90 \\
HD201891          & 24.3$\pm$0.5 & HIRES       & 0.000$\pm$0.000 &nearby    & 5947 & 4.31 & -1.05 & 0.788$\pm$0.010 & 2.01 & 1.96 \\
HD205650          & 15.6$\pm$0.5 & UVES (NS10) & 0.000$\pm$0.000 &NaD+nearby& 5952 & 4.47 & -1.16 & 0.779$\pm$0.013 & 1.82 & 1.78 \\
HD213657          & 30.0$\pm$0.5 & A06         & 0.002$\pm$0.002 &NaD       & 6299 & 3.90 & -1.96 & 0.795$\pm$0.023 & 2.33 & 2.26 \\
HD218502          & 28.6$\pm$0.8 & UVES archive& 0.000$\pm$0.005 &NaD+nearby& 6298 & 3.96 & -1.76 & 0.791$\pm$0.010 & 2.31 & 2.24 \\
HD219617          & 37.8$\pm$0.8 &UVES arc+MR04& 0.001$\pm$0.002 &NaD       & 6146 & 4.10 & -1.48 & 0.790$\pm$0.012 & 2.37 & 2.30 \\
HD233511          & 31.9$\pm$1.5 & FIES/NOT    & 0.002$\pm$0.002 &NaD       & 6127 & 4.29 & -1.63 & 0.759$\pm$0.010 & 2.25 & 2.19 \\
HD241253          & 19.8$\pm$0.4 & UVES (NS10) & 0.001$\pm$0.001 &NaD       & 6023 & 4.37 & -1.05 & 0.818$\pm$0.010 & 1.98 & 1.93 \\
HD250792          & 16.0$\pm$0.7 & FIES/NOT    & 0.002$\pm$0.002 &NaD+nearby& 5568 & 4.40 & -1.01 & 0.715$\pm$0.010 & 1.50 & 1.50 \\
HD284248          & 32.0$\pm$1.0 & UVES archive& 0.020$\pm$0.005 &NaD       & 6285 & 4.29 & -1.56 & 0.810$\pm$0.010 & 2.37 & 2.29 \\
HD298986          & 30.4$\pm$0.5 & A06         & 0.004$\pm$0.002 &NaD       & 6223 & 4.26 & -1.34 & 0.825$\pm$0.010 & 2.30 & 2.23 \\
HD338529          & 23.5$\pm$0.5 & A06+HIRES   & 0.008$\pm$0.002 &NaD       & 6426 & 3.89 & -2.29 & 0.795$\pm$0.018 & 2.30 & 2.23 \\
LP0635-0014       & 22.3$\pm$0.5 &  HIRES      & 0.060$\pm$0.010 &NaD       & 6516 & 3.98 & -2.50 & 0.808$\pm$0.041 & 2.35 & 2.28 \\
LP0815-0043       & 18.9$\pm$0.5 & A06         & 0.024$\pm$0.015 &M06       & 6534 & 4.17 & -2.78 & 0.781$\pm$0.010 & 2.26 & 2.20 \\
LP0831-0070       & 24.3$\pm$0.6 & UVES archive& 0.005$\pm$0.002 &NaD       & 6414 & 4.49 & -2.94 & 0.778$\pm$0.029 & 2.28 & 2.21 \\
BPS BS16023-0046  & 19.3$\pm$0.6 & UVES (B07)  & 0.004$\pm$0.010 &maps      & 6547 & 4.50 & -2.90 & 0.811$\pm$0.027 & 2.27 & 2.20 \\
BPS BS16968-0061  & 27.9$\pm$0.3 & UVES (B07)  & 0.019$\pm$0.010 &maps      & 6256 & 3.88 & -3.07 & 0.823$\pm$0.044 & 2.23 & 2.18 \\
BPS BS17570-0063  & 17.6$\pm$0.4 & UVES (B07)  & 0.026$\pm$0.010 &maps      & 6318 & 4.69 & -2.91 & 0.76 $\pm$0.020 & 2.06 & 1.99 \\
BPS BS17572-0100  & 18.5$\pm$0.9 & UVES archive& 0.016$\pm$0.002 &NaD       & 6596 & 4.32 & -2.66 & 0.802$\pm$0.022 & 2.28 & 2.22 \\
BPS CS22177-0009  & 24.2$\pm$0.3 & UVES (B07)  & 0.023$\pm$0.010 &maps      & 6421 & 4.57 & -3.04 & 0.786$\pm$0.027 & 2.28 & 2.21 \\
BPS CS22888-0031  & 18.7$\pm$0.4 & UVES (B07)  & 0.007$\pm$0.008 &maps      & 6335 & 4.99 & -3.24 & 0.76 $\pm$0.020 & 2.11 & 2.04 \\
BPS CS22953-0037  & 19.5$\pm$0.3 & UVES (B07)  & 0.008$\pm$0.010 &maps      & 6532 & 4.33 & -2.84 & 0.785$\pm$0.018 & 2.27 & 2.21 \\
BPS CS22966-0011  & 13.7$\pm$0.5 & UVES (B07)  & 0.000$\pm$0.010 &maps      & 6307 & 4.73 & -3.06 & 0.75 $\pm$0.020 & 1.92 & 1.88 \\
BPS CS29491-0084  & 18.0$\pm$0.7 & UVES (B07)  & 0.003$\pm$0.005 &NaD+maps  & 6393 & 4.07 & -2.94 & 0.767$\pm$0.017 & 2.14 & 2.09 \\
BPS CS29518-0020  & 21.0$\pm$1.1 & UVES (B07)  & 0.009$\pm$0.010 &maps      & 6464 & 4.67 & -2.72 & 0.76 $\pm$0.020 & 2.25 & 2.19 \\
BPS CS29518-0043  & 17.2$\pm$1.1 & UVES (B07)  & 0.008$\pm$0.010 &maps      & 6517 & 4.28 & -3.17 & 0.775$\pm$0.019 & 2.20 & 2.16 \\
BPS CS29527-0015  & 18.6$\pm$0.6 & UVES (B07)  & 0.014$\pm$0.007 &maps      & 6541 & 4.24 & -3.43 & 0.776$\pm$0.015 & 2.25 & 2.19 \\
BPS CS31061-0032  & 21.0$\pm$0.6 & UVES (B07)  & 0.015$\pm$0.010 &maps      & 6433 & 4.32 & -2.57 & 0.770$\pm$0.017 & 2.23 & 2.18 \\
\hline           
\end{tabular}
\end{table}


\end{document}